\documentclass[prd,showpacs,showkeys,nofootinbib,superscriptaddress]{revtex4}
\usepackage{graphicx,amssymb}

\begin{document}

\title{A Conserved Vector Current test using low energy beta-beams}

\author{A.~B.~Balantekin}
\email{baha@physics.wisc.edu}
\affiliation{Department of Physics, University of Wisconsin, Madison, WI 53706, USA}
\author{J.~H.~de~Jesus}
\email{jhjesus@physics.wisc.edu}
\affiliation{Department of Physics, University of Wisconsin, Madison, WI 53706, USA}
\author{R.~Lazauskas}
\email{lazauska@lpsc.in2p3.fr}
\affiliation{Institut de Physique Nucl\'eaire, F-91406 Orsay cedex, France}
\author{C.~Volpe}
\email{volpe@ipno.in2p3.fr}
\affiliation{Institut de Physique Nucl\'eaire, F-91406 Orsay cedex, France}
\date{\today }

\begin{abstract}
We discuss the possibility of testing the weak currents and, in particular, the weak magnetism term through the measurement of the $\overline{\nu}_e + p \rightarrow e^+ + n$ reaction at a low energy beta-beam facility.  We analyze the sensitivity using both the total number of events and the angular distribution of the positrons emitted in a water \v{C}erenkov detector.  We show that the weak magnetism form factor might be determined with better than several percent accuracy using the angular distribution.  This offers a new way of testing the Conserved Vector Current hypothesis. 
\end{abstract}

\keywords{Weak magnetism, neutrino-nucleon reactions, beta-beams}
\pacs{13.15.+g, 23.40.Bw, 25.30.Pt}
\maketitle

\section{Introduction}
Cross sections for neutrino-nucleon reactions were among the first predictions of the Standard Model of electroweak interactions~\cite{Weinberg:1972tu}.  In addition to being one of the first test cases of the Standard Model, these reactions play a crucial role in understanding the dynamics of the core-collapse supernovae~\cite{Balantekin:2003ip}.  Neutrino interactions control not only the dynamics of core-collapse supernovae, but also the yields of the r-process nucleosynthesis that could take place in such environments~\cite{Meyer:1998sn}.  Neutrino-nucleon interactions also contribute to the energy transfer from the accretion-disk neutrinos to the nucleons in models of gamma-ray bursts~\cite{Ruffert:1996by,Kneller:2004jr}.  Finally, understanding the subtleties of the neutrino-nucleon interactions are crucial to the terrestrial observation of neutrino signals~\cite{Vogel:1999zy,Beacom:2002hs}.

Even though there are extensive measurements of the neutrino-nucleon scattering at GeV energies~\cite{Ahrens:1986xe}, very little data exist in the energy range up to 100 MeV, relevant to the astrophysical applications.  Neutrino-nucleus scattering experiments in this energy range can be carried out either at intense, pulsed spallation sources~\cite{Avignone:2003ep,Scholberg:2005qs}, or at low energy beta-beam facilities~\cite{Volpe:2003fi}.  Beta-beams are pure $\nu_e$ or $\overline{\nu}_e$ beams produced by allowing radioactive ions circulating in a storage ring to decay~\cite{Zucchelli:2002sa,bbeams:0000}.  Low energy beta-beams, where one can vary the Lorentz boost factor $\gamma$ of the stored ions, offer a particular advantage to extract information about the energy dependence of the cross section in the energy range up to 100 MeV.  So far, several applications have been discussed in the literature concerning neutrino-nucleus scattering~\cite{Serreau:2004kx,McLaughlin:2004va,Volpe:2005iy}, electroweak tests of the Standard Model~\cite{McLaughlin:2003yg,Balantekin:2005md}, as well as core-collapse supernova physics~\cite{Volpe:2003fi,Jachowicz:2005ym}.

In this paper, we explore the possibility of performing a Conserved Vector Current (CVC) test by measuring the interaction of anti-neutrinos on protons in a water \v{C}erenkov detector, using low energy beta-beams.  The conserved vector current hypothesis connects weak and electromagnetic hadronic currents.  Several tests of CVC have been performed in the past (see e.g. Ref.~\cite{Thomas:1996ap}).  In particular, the CVC prediction on the vector form factor has been extensively studied in super-allowed nuclear beta-decays (see e.g. Ref.~\cite{Hardy:2004id}).  Verifying that the CVC hypothesis correctly predicts tensor terms (often referred to as weak magnetism) is of fundamental importance~\cite{Deutsch:1977uv}.  The CVC prediction of the weak magnetism has been tested in an experiment involving the beta-decay of Gamow-Teller transitions in mirror nuclei in the A=12 triad~\cite{Lee:1963,Wu:1964,Lee:1965}.  One should note that whenever the allowed contributions to certain Gamow-Teller transitions are inhibited due to accidental cancellations, weak magnetism terms may contribute appreciably~\cite{Towner:2005qc}.  Here, we propose a test based on neutrino-nucleon collisions at low momentum which would, in particular, have the advantage that there is no uncertainty coming from nuclear structure calculations.  In addition to providing a test of the CVC hypothesis, measurement of weak magnetism terms in neutrino-nucleon interactions are of direct interest to astrophysics as these terms may play an important role in the dynamic of core-collapse supernova~\cite{Horowitz:2001xf}.

The structure of the paper is as follows. In Section II, we describe the first- and second-class weak currents.  Section III presents the formalism used in this work, concerning the scattering of anti-neutrinos on protons, as well as its angular distribution.  The results are given in Section IV, while the conclusions are made in Section V.

\section{First- and second-class weak currents}
The most general covariant form for the matrix element of the reaction $\overline{\nu }_{e}+p\rightarrow e^{+}+n$ is~\cite{Weinberg:1958ut,Azimov:1961,Eisenberg:1988,Towner:1995}
\begin{eqnarray}
\label{eqn:weak_curr}
{\mathcal{M}}~=~\frac{G_{F}\cos \theta _{C}}{\sqrt{2}}\left\{ \overline{u}_{n}\left[ \gamma _{\alpha }(f_{1}-g_{1}\gamma _{5})+\sigma _{\alpha \beta}k^{\beta }(f_{2}+g_{2}\gamma _{5})+k_{\alpha }(f_{3}+g_{3}\gamma _{5})\right] u_{p}\right\} \left\{ \overline{\nu }_{\nu }\gamma^{\alpha }(1-\gamma _{5})\nu _{e}\right\}~,
\end{eqnarray}
where $G_{F}$ is the Fermi weak coupling constant, $\cos \theta_{C}=0.974$ is the Cabibbo angle, and the $f$'s and $g$'s are invariant form factors that depend on the transferred momentum $q^{2}\equiv (p_{p}-p_{n})^{2}$.  The different terms correspond to the vector $f_{1}$, the axial-vector $g_{1}$, the tensor $f_{2}$ (or weak magnetism), the induced tensor $g_{2}$, the induced scalar $f_{3}$ and the induced pseudo-scalar $g_{3}$.  These currents are classified according to the \cal{G}-parity transformation: the first-class currents ($f_{1},$ $g_{1},$ $f_{2}$ and $g_{3}$) are the \cal{G}-parity invariant ones, while the second-class currents ($g_{2}$ and $f_{3}$) are related to \cal{G}-parity breaking Lagrangian terms.

Under the postulate~\cite{Towner:1995,Feynman:1958} that the weak vector and isovector electromagnetic currents are members of an isotriplet vector of current operators, the CVC hypothesis states that
\begin{eqnarray}
\label{eqn:ffac0}
\lim_{q^2 \rightarrow 0} f_1(q^2) ~=~ 1~; \hspace{0.75cm} \lim_{q^2 \rightarrow 0} f_2(q^2) ~=~ \frac{\mu_p-\mu_n}{2m_N}~; \hspace{0.75cm} f_3(q^2) ~=~ 0~,
\end{eqnarray}
where $\mu _{p}-\mu _{n}=3.706$ is the difference in the anomalous magnetic moments of the nucleons, and $m_{N}$ is the mass of the nucleon.  It is usually assumed that the momentum dependence of the $f$ and $g$ form factors are of the form
\begin{eqnarray}
\label{eqn:ffac1}
f_1(q^2) ~=~ \left[1+\frac{q^2}{\left(0.84~\mathrm{GeV}\right)^2}\right]^{-2}~; \hspace{0.75cm} f_2(q^2) ~=~ \left(\frac{\mu_p-\mu_n}{2m_N}\right) f_1(q^2)~; \nonumber \\
g_1(q^2) ~=~ -1.262\left[1+\frac{q^2}{\left(1.032~\mathrm{GeV}\right)^2}\right]^{-2}~; \hspace{0.75cm} g_3(q^2) ~=~ \left(\frac{2m_N}{q^2+m_\pi^2}\right) g_1(q^2)~,
\end{eqnarray}
with $m_{\pi }$ being the mass of the pion. The weak magnetism $f_2$ and vector $f_1$ form factors, and the pseudo-scalar $g_3$ and axial-vector $g_1$ form factors, are connected via the Goldberger-Trieman relation~\cite{Golberger:1958}.  Little is known about the existence or structure of the induced tensor $g_2$, which together with the induced scalar $f_3$ have been the object of a longstanding search~\cite{Wilkinson:2000gx}.

The detailed expression of the square of the matrix element of Eq.~(\ref{eqn:weak_curr}) is given in the Appendix.  Vector and axial-vector contributions to weak processes have been tested in various processes, e.g. in super-allowed nuclear $\beta$ decays~\cite{Hardy:2004id}.  The contributions to it from the induced scalar $f_3$ and from the induced pseudo-scalar $g_3$, turn out to be proportional to the square of the emitted lepton mass (positron, in our case).  This considerably suppresses their impact\footnote{The induced scalar and induced pseudo-scalar contribution are enhanced by a factor of $\sim 10^{4}$ and become sizable in the processes involving heavy leptons $\overline{\nu }_{\mu }+p\rightarrow \mu ^{+}+n~$, $\overline{\nu}_{\mu }+p\rightarrow \mu ^{+}+\Lambda ~$, and $\overline{\nu}_{\mu }+p\rightarrow \mu ^{+}+\Sigma ^{0}$, with thresholds in the laboratory frame of 113~MeV, 319~MeV and 420~MeV, respectively.  However, for the cross section of the process $\overline{\nu}_e + p \rightarrow e^+ + n$ below 100 MeV, these terms contribute less than $10^{-4}$\%.} relative to the other four form factors and therefore will not be considered further.  The same is not true regarding the weak magnetism $f_2$ or induced tensor $g_2$ contributions.  These two terms enter in similar ways in the anti-neutrino proton cross section -- see Eq.~(\ref{eqn:m2}) -- and their contribution will be proportional to the corresponding strengths $f_2$ and $g_2$.  Therefore it would be very hard to disentangle the contributions of these two terms in any weak process.  This means that a possible deviation of the weak magnetism contribution from the CVC prediction~(\ref{eqn:ffac1}), can equally be attributed to a non-zero $g_2$ second-class current.

In low energy beta-beam experiments, the (anti)neutrinos have energies ranging from a few eV's to 100 MeV; thus, the relevant transferred momentum $q$ is small compared to the hadron mass.  As a result, in these experiments, the weak magnetism dependence on $q^2$ does not have any sizable effect.  Without losing accuracy, we assume
\begin{eqnarray}
\label{eqn:wmCVC}
f_2 ~\equiv~ f_{2}(0) ~=~ \frac{\mu _{p}-\mu _{n}}{2m_{N}}~,
\end{eqnarray}
which is the zero-momentum transfer limit of the CVC hypothesis~(\ref{eqn:ffac0}).

Under these assumptions, a low energy beta-beam experiment might be able to probe  $f_{2}$ with a precision determined by its level of statistical and systematic errors, and provide a test of CVC.  Note that it is possible to carry out tests of CVC in high-energy neutrino reactions as well~\cite{Adler:1964yx}.

\section{Angular distribution and number of events}
At a beta-beam facility, the (anti)neutrino flux is produced by boosted radioactive ions decaying in a storage ring~\cite{Zucchelli:2002sa}. In Ref.~\cite{Serreau:2004kx}, it was shown that a small storage ring is more appropriate for low energy beta-beam experiments. The feasibility of such a storage ring is now ongoing~\cite{Chance:0000}.  With the aim of extracting the strength of the weak magnetism form factor~(\ref{eqn:wmCVC}), we focus on anti-neutrino capture by protons in a water \v{C}erenkov detector.  Using such setup, both the total number of events and the angular distribution of the emitted positrons can be measured.

\begin{figure}[t]
\includegraphics[width=7cm]{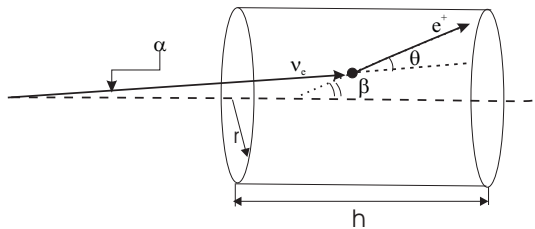}
\caption{Schematic view of the geometry of a $\overline{\protect\nu}_e + p \rightarrow e^+ + n$ event in a water detector.}
\label{fig:exp_pic}
\vskip 0.5cm
\end{figure}

The differential cross section for the anti-neutrino capture by protons is
\begin{eqnarray}
\label{eqn:cross}
\frac{d\sigma}{dt} ~=~ -\frac{G_F^2\cos\theta_C}{4\pi(s-m_p^2)^2}\,(1+\Delta_R)\left\vert \mathcal{M}\right\vert^2~,
\end{eqnarray}
where $\Delta _{R}\simeq 0.024$~\cite{Wilkinson:1994ua,Hardy:1998eu} accounts for the inner energy-independent radiative corrections.  The Mandelstam variables $s$ and $t\equiv q^{2}$, as well as an expression of $\left\vert \mathcal{M}\right\vert ^{2}$, are provided in the Appendix.  It is straightforward to relate the cross section in Eq.~(\ref{eqn:cross}) to the differential cross section of the emitted positrons, by multiplying it by a factor $dt/d(\cos\theta)$ (see the Appendix).  In order to obtain the observed angular distribution, one has to consider both the convolution of the neutrino flux and the geometry of the detector (Fig.~\ref{fig:exp_pic}).  In particular, one has to integrate over the useful part of the storage ring, the volume of the detector, and the entire energy spectrum of the emitted neutrinos
\begin{eqnarray}
\label{eqn:crossangle}
\frac{dN}{d\cos \beta} ~=~ f\tau nh\Delta t\int_{0}^{\infty}dE_{\nu}\int_{0}^{D}\frac{d\ell }
{L}\int_{0}^{h}\frac{dz}{h}\int_{-\alpha _{\max }}^{\alpha _{\max }}\frac{\sin \alpha d\alpha }{4}~
\Phi _{lab}(E_{\nu },\ell,z,\alpha )\left( \frac{d\sigma }{d\cos \theta}\right) \left( \frac{d\cos \theta }{d\cos \beta }\right)~,
\end{eqnarray}
where $f$ is the number of injected ions per unit time, $\tau=t_{1/2}/\ln 2$ is the lifetime of the parent nuclei, $n$ is the number of protons per unit volume, $h$ is the depth of the detector, $L$ is the length of the storage ring straight sections, and $\Delta t$ is the duration of the measurement.  The detailed expression for the neutrino flux in the laboratory frame $\Phi _{lab}$ can be found in Ref.~\cite{Serreau:2004kx}.  Note the different angles that enter Eq.~(\ref{eqn:crossangle}): $\alpha$ is the angle at which the neutrino hits the detector; the angles $\theta$ and $\beta$ are the positron emission angles relative to the neutrino and to the detector axis, respectively (see Fig.\ref{fig:exp_pic}).  The total number of events observed in the detector is obtained by integrating the last expression over $\cos\beta$.

The angular dependence of the functional defined in Eq.~(\ref{eqn:crossangle}) can be expanded in a Legendre polynomial $P_{n}(\cos \beta )$ basis, using the following integral relations
\begin{eqnarray}
\label{eqn:development}
\frac{dN}{d\cos \beta} ~=~ \sum_{n=0}^{\infty }A_{n}P_{n}(\cos\beta)~,
\end{eqnarray}
and
\begin{eqnarray}
\label{eqn:coefficients}
A_{n} ~=~ \frac{2n+1}{2}\int_{-1}^{1}P_{n}(\cos \beta)\left(\frac{dN}{d\cos \beta}\right)d\cos \beta~.
\end{eqnarray}
Note that the zeroth order expansion coefficient, $A_{0}$, is nothing but the total number of events. The first order coefficient, $A_{1}$, measures the strength of a linear term in the angular distribution which favors forward peaked events.

\section{Results and discussion}
In the case of low energy beta-beams~\cite{Volpe:2003fi}, the Lorentz ion boosts are in the range of 7 to 14.  In our calculations, we assume a fully-efficient $954\times 10^{3}$~kg water \v{C}erenkov detector located at $d=10$~m from the storage ring, which has $L=1885$~m total length and $D=678$~m straight sections~\cite{Chance:0000}.  The number of stored $^{6}$He ions producing the anti-neutrinos is $f_{^{6}\mathrm{He}}=2.7\times 10^{12}$~ions/s, while the intensity of $^{6}$He ions at production is $3 \times 10^{13}$~ions/s.  Since the feasibility study is still ongoing, the characteristics of the storage ring and of the setup should be considered as preliminary~\cite{Chance:0000}.  This configuration was used in Ref.~\cite{Balantekin:2005md} to study a possible measurement of the Weinberg angle, through elastic scattering of (anti)neutrinos on electrons.

\begin{figure}[t]
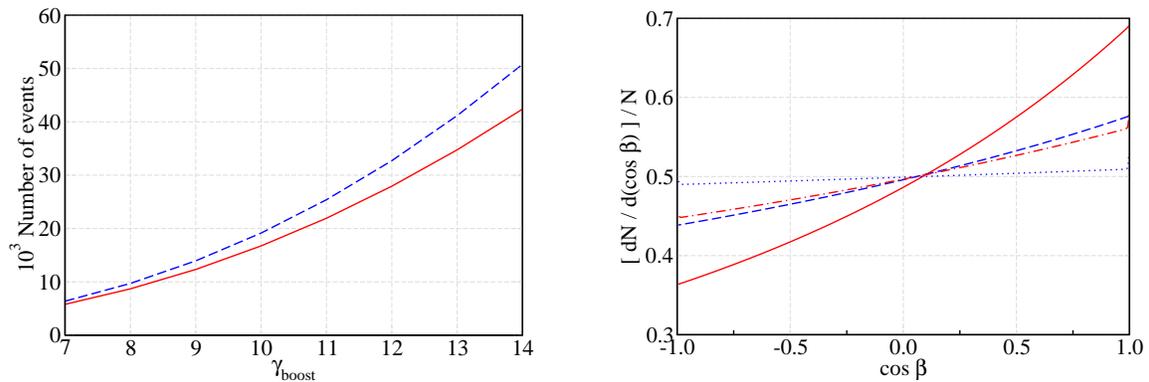

\includegraphics[width=7cm]{fig2a.eps}
\hskip 1cm
\includegraphics[width=7cm]{fig2b.eps}
\caption{Left-hand panel: Expected number of events at the detector over a period of one year ($3\times 10^{7}$~s) as a function of the Lorentz ion boost $\gamma$, with (solid) and without (dashed) the weak magnetism term.  Note that at the highest $\gamma$, the weak magnetism suppresses the total number of events by as much as 17\%.  Right-hand panel: Angular distribution of the positrons emitted in anti-neutrino capture by protons as observed in the water detector -- Eq.~(\ref{eqn:crossangle}) -- normalized by the total number of events.  The four different curves correspond to two extreme values of the Lorentz boost, with (solid is for $\gamma=14$; dash-dotted is for $\gamma=7$) and without (dashed is for $\gamma=14$; dotted is for $\gamma=7$) the weak magnetism term.}
\label{fig:nevents}
\vskip 0.5cm
\end{figure}

On the left-hand panel of Fig.~\ref{fig:nevents}, we present a plot with the number of events expected at the detector over a period of one year ($\Delta t=3\times 10^{7}$~s) for different Lorentz boosts of the ions ($\gamma =7-14$).  One can see that the weak magnetism form factor of Eq.~(\ref{eqn:wmCVC}) tends to reduce the anti-neutrino on proton cross section: it is responsible for suppressing the expected number of events by approximately 17\% for the largest Lorentz boost we consider ($\gamma =14$); for the lowest boost, the suppression of events is 10\%.  In addition, a strong enhancement of the total number of counts is achieved when increasing the Lorentz boost: for $\gamma =14$, one expects seven times larger statistics than for $\gamma =7$.  Both these features indicate that performing an experiment with a gamma value as high as possible is advantageous.

\begin{figure}[t]
\includegraphics[width=7cm]{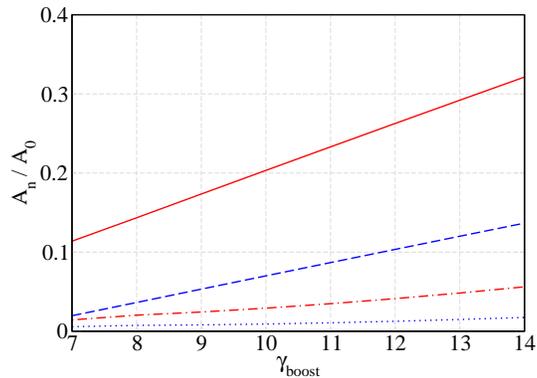}
\caption{Ratio of the first ($A_1$) and second ($A_2$) order Legendre polynomials to the zeroth order one ($A_0$, the number of events) with (solid is for $A_1/A_0$; dash-dotted is for $A_2/A_0$) and without (dashed is for $A_1/A_0$; dotted is for $A_2/A_0$) the weak magnetism term, as a function of the Lorentz boost.  The overall effect of the weak magnetism term on the angular distribution is considerably larger than on the total number of events. Note that when the weak magnetism is considered, its value is set to the one of Eq.~(\ref{eqn:wmCVC}).}
\label{fig:coefficients}
\vskip 0.5cm
\end{figure}

On the right-hand panel of Fig.~\ref{fig:nevents}, we show how the angular distribution is modified by the weak magnetism term for the two extreme values of the Lorentz boost, $\gamma=7$ and $\gamma=14$.  One can see that this term tends to favor forward peaked positrons events and suppress backward ones.  As a consequence, the weak magnetism term leaves a strong signature on the asymmetry of the angular distribution of the emitted positrons.  As discussed in the previous section, this asymmetry can be studied by performing an expansion of the positron angular distribution function into Legendre polynomials -- Eqs.~(\ref{eqn:development}) and~(\ref{eqn:coefficients}).  This distribution turns out to be smooth and almost linear. Therefore, only a few lower order Legendre polynomials -- and, in particular, the $A_{1}$ term, which is related to a linear increase of the forward peaked events -- are relevant in parameterizing it.  The variation of these coefficients as a function of the Lorentz boosts is shown in Fig.~\ref{fig:coefficients}, both when the value~(\ref{eqn:wmCVC}) for the weak magnetism is considered, and when this term is switched off.  The overall effect of this term on the angular distribution is considerably larger than on the total number of events (Fig.~\ref{fig:nevents}).  This indicates that, in principle, the angular distribution of events is a better tool to measure the weak magnetism.  Furthermore, the asymmetry of the angular distribution is a dimensionless quantity and should not suffer from possible normalization problems, eventually having smaller systematic errors.

\begin{figure}[t]
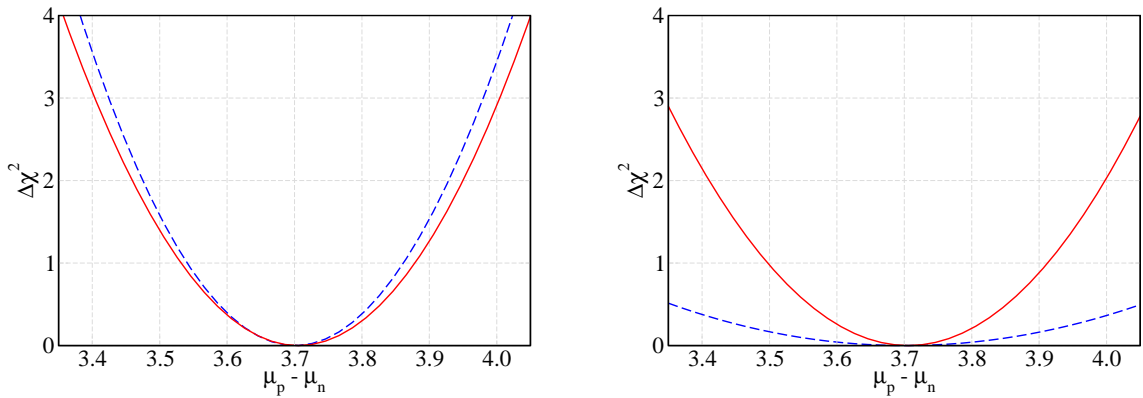

\includegraphics[width=7cm]{fig4a.eps}
\hskip 1cm
\includegraphics[width=7cm]{fig4b.eps}
\caption{$\Delta \chi^2$ obtained from the total number of events (dashed) and from the angular distribution of events (solid) in the cases when the total error is purely statistical (left-hand panel) and when a 2\% systematic error is added to the total error (right-hand panel).  For a purely statistical error, the 1$\sigma$ ($\Delta \protect\chi^2$=1) relative uncertainty in $\mu_p - \mu_n$ is close to 5\% from both the number of events and the angular distribution analysis.  Even though the angular distribution is more sensitive to the weak magnetism term, its smaller statistical error makes the analysis through the number of events more advantageous in the absence of systematic errors.  As soon as a $2\%$ systematic error is added to the total error in both the number of events and in the angular distribution, the latter becomes a better method to measure the weak magnetism.  These results were obtained considering $\gamma=12$.}
\label{fig:chi2}
\vskip 0.5cm
\end{figure}

The level at which one can measure the weak magnetism form factor at a low energy beta-beam facility crucially depends on its statistical and systematic errors.  We address this problem by taking as an example an experiment running for one year ($\Delta t=3\times 10^{7}$~s), with radioactive ions boosted to $\gamma=12$.  If one assumes no systematic error, one could extract the value of the weak magnetism form factor within 5\% accuracy (see the left-hand panel of Fig.~\ref{fig:chi2}).  Furthermore, one can see that, in this particular case, it is slightly advantageous to consider the total number of events instead of the angular distribution.  This is due to the fact that the statistical error slightly increases when one separates the total number of events into angular bins to extract the asymmetry factor of the angular distribution\footnote{Note that the statistical error in the number of events is less than 1\%, while the statistical error for the angular distribution, obtained by considering that these events are divided into 18 bins, is almost 4\%.}.  However, as soon as a $2$\% systematic error is included, the level of accuracy on the weak magnetism term from the total number of events is substancially reduced (right-hand panel of Fig.~\ref{fig:chi2}), making this procedure inviable due to the low sensitivity of this observable to those terms.  Instead, at this level of systematic error, the angular distribution becomes a better way to extract information on the weak magnetism form factor.

\begin{figure}[t]
\includegraphics[width=7cm]{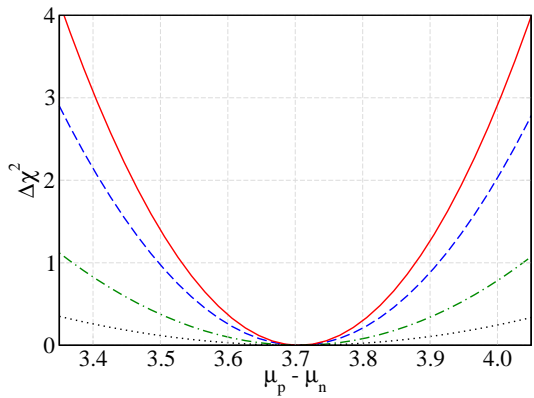}
\caption{$\Delta \chi^2$ obtained from the angular distribution of events in the cases when the total error is purely statistical (solid), and when one includes 2\% (dashed), 5\% (dash-dotted) and 10\% (dotted) systematic errors.  The 1$\sigma$ ($\Delta \chi^2=1$) relative uncertainty in $\mu_p - \mu_n$ is 4.7\%, 5.6\%, 9.0\% and more than 20\%, respectively.  These results were obtained considering $\gamma=12$.}
\label{fig:chi2sys}
\vskip 0.5cm
\end{figure}

In general, systematic errors are expected to be more important than statistical ones due to possible backgrounds (e.g. beam or cosmic-ray induced).  In Fig.~\ref{fig:chi2sys}, we show the $\Delta \chi^2$ distribution for different levels of systematic errors in the analysis of the angular distribution of events.  When 2\%, 5\% and 10\% systematic errors are added to the statistical error, the 1$\sigma$ ($\Delta \chi^2=1$) relative uncertainty in $\mu_p - \mu_n$ is 5.6\%, 9.0\% and more than 20\%, respectively.  It is clear that systematic errors have a very strong impact on how well we can extract information about the weak magnetism form factor~(\ref{eqn:wmCVC}) at low energy beta-beam facilities.

\section{Conclusions}
In this work, we discuss the possibility of measuring the weak magnetism term of the weak currents.  The method we propose exploits anti-neutrino capture by protons, i.e. $\overline{ \nu}_e + p\rightarrow e^+ + n$, in a water \v{C}erenkov detector.  The anti-neutrinos are assumed to be produced at a low energy beta-beam facility.  We study the sensitivity that can be achieved both with the total number of events and with the angular distribution of the emitted positrons.

Our calculations show that, when systematic errors are taken into account, the angular distribution is a much better tool than the total number of events. If those errors are kept below 5\%, a one year measurement of the weak magnetism term is possible at a $1\sigma$ level of 9\%, if the ions in the storage ring are boosted to $\gamma=12$.  A feasibility study for the low energy beta-beam facility is now ongoing.  In this work, we use the predicted $^6$He intensity from a preliminary study.  Because of the increasing importance of the weak magnetism contribution with the impinging neutrino energy, a better measurement is expected if the ions are boosted to $\gamma$'s larger than 12.  This way of probing the weak magnetism form factor at low momentum constitutes a new test of the Conserved Vector Current hypothesis.

\begin{acknowledgments}
We would like to thank J.~Hardy for useful discussions.  The authors acknowledge the CNRS-Etats Units 2005 and 2006 grants which have been used during the completion of this work.  This work was also supported in part by the EURISOL design study, in part by the U.S. National Science Foundation Grant No. PHY-0244384 at the University of Wisconsin, and in part by the University of Wisconsin Research Committee with funds granted by the Wisconsin Alumni Research Foundation.
\end{acknowledgments}

\appendix*
\section{}
For the sake of clarity, we give the differential cross section for the reaction $\overline{\nu}_e + p \rightarrow e^+ +n$~\cite{Azimov:1961}
\begin{eqnarray}
\label{eqn:diffsigma2}
\frac{d\sigma}{dt}~=~-\frac{G_F^2\cos \theta _C}{4\pi (s-m_p^2)^2}\,(1+\Delta_R)\left\vert \mathcal{M}\right\vert ^2~.
\end{eqnarray}
where $\Delta_R \simeq 0.024$~\cite{Wilkinson:1994ua,Hardy:1998eu}.  The Mandelstam variables are defined as
\begin{eqnarray}
\label{eqn:mandel}
s &=& (p_p+p_\nu)^2~=~m_p^2+2m_pE_\nu~, \nonumber \\
t &\equiv&q^2~=~(p_e-p_\nu)^2~=~m_n^2-m_p^2-2m_p(E_\nu-E_e)~=~m_e^2-2E_\nu E_e+2E_\nu \sqrt{E_e^2-m_e^2} \cos\theta~, \\
u &=& (p_p-p_e)^2~=~m_p^2+m_e^2-2m_pE_e~, \nonumber
\end{eqnarray}
and the square of the reaction matrix element~(\ref{eqn:weak_curr}) is~\cite{Azimov:1961}
\begin{eqnarray}
\label{eqn:m2}
\left\vert \mathcal{M} \right\vert^2 &=& \left\vert f_1-g_1 \right\vert^2\left(m_n^2-u\right) \left(m_e^2+m_p^2-u\right)  \nonumber \\
&+& \left\vert f_1+g_1 \right\vert^2 \left(s-m_p^2\right)\left(s-m_e^2-m_n^2\right)  \nonumber \\
&+& \left(\left\vert f_2 \right\vert^2 + \left\vert g_2 \right\vert^2\right) \times  \nonumber \\
&& \times \left\{ t\left[\left(s-m_p^2\right) \left(u-m_n^2\right) + \left(s-m_n^2\right) \left(u-m_p^2\right) \right] - m_e^2 \left[\left(m_n^2-m_p^2\right) \left(s-u\right) + \frac{1}{2} \left(m_e^2-t\right)\left(m_n^2+m_p^2+t\right) \right] \right\}  \nonumber \\
&+& 2\,{\rm Re} \left(f_2g_2^\ast\right) \left(m_n^2-m_p^2\right) \left[t\left(s-u\right)+m_e^2\left(m_n^2-m_p^2\right)\right] \nonumber \\
&+& \frac{1}{2} \left(\left\vert f_3\right\vert^2+\left\vert g_3\right\vert^2\right) m_e^2\left(m_e^2-t\right) \left(m_p^2+m_n^2-t\right) \\
&+& m_pm_n \left(m_e^2-t\right) \left[ 2\left(\left\vert g_1\right\vert^2-\left\vert f_1\right\vert^2\right) + \left(2t+m_e^2\right) \left(\left\vert g_2\right\vert^2-\left\vert f_2\right\vert^2\right) -m_e^2 \left(\left\vert g_3\right\vert^2-\left\vert f_3\right\vert^2\right) \right] \nonumber \\
&+& 2\,{\rm Re} \left[f_1g_2^\ast \left(m_n-m_p\right)-g_1f_2^\ast\left(m_n+m_p\right) + \frac{1}{2} \left(f_2f_3^\ast + g_2g_3^\ast\right)m_e^2\right] \left[t\left(s-u\right)+m_e^2\left(m_n^2-m_p^2\right)\right]  \nonumber \\
&+& 2\,{\rm Re} \left(f_1f_2^\ast+g_1g_2^\ast\right) m_p\left[t\left(t-m_p^2+m_n^2\right)+m_e^2\left(s-m_n^2-m_e^2\right)\right] \nonumber \\
&+& 2\,{\rm Re} \left(f_1f_2^\ast-g_1g_2^\ast\right) m_n\left[t\left(t-m_n^2+m_p^2\right)+m_e^2\left(u-m_p^2-m_e^2\right)\right] \nonumber \\
&+& 2\,{\rm Re} \left(f_1f_3^\ast+g_1g_3^\ast\right) m_e^2m_p \left(m_n^2-t\right)\nonumber \\
&+& 2\,{\rm Re} \left(f_1f_3^\ast-g_1g_3^\ast\right) m_e^2m_n\left(s-m_p^2\right)~. \nonumber
\end{eqnarray}
To get an expression for the differential cross section~(\ref{eqn:diffsigma2}) depending on the scattering angle, one uses
\begin{eqnarray}
\label{eqn:diffangle}
\frac{dt}{d\cos \theta} ~=~ \frac{2E_\nu E_e m_p}{m_p + E_\nu -E_\nu \frac{E_e}{\sqrt{E_e^2-m_e^2}}\cos \theta}~.
\end{eqnarray}
\noindent
From the neutrino energy in the laboratory frame $E_\nu$, we define the parameter $\eta =m_p/(m_p+E_\nu)$. Then, the energy of the positrons in the center of mass frame is
\begin{eqnarray}
\label{eqn:een}
E_{e;cm} ~=~ \sqrt{\frac{m_p^2 \left(1+\eta\right)^2 -\left(m_n^2-m_e^2\right) 
\left(1-\eta^2\right)} {2m_p\left(1+\eta\right)\sqrt{1-\eta^2}}}~.
\end{eqnarray}
For maximal and minimal square of the positrons energy in the laboratory frame, one has the relations
\begin{eqnarray}
\label{eqn:minmax}
E_e^2 ~=~ \frac{E_{e;cm}^2 \left(1+\eta^2\right) + m_e^2\eta^2 \pm 2\eta
E_{e;cm} \sqrt{E_{e;cm}^2 - m_e^2}}{1-\eta^2}~.
\end{eqnarray}

\end{document}